\documentclass[aps,prb,twocolumn,amsmath,amssymb,nofootinbib,superscriptaddress,floatfix,citeautoscript,showpacs]{revtex4-1}

\usepackage[colorlinks=true,citecolor=blue,urlcolor=blue,linkcolor=blue]{hyperref}
\usepackage{bbm}
\usepackage{graphicx}
\usepackage{bm}

\newcommand{\zz}{\mathbb{Z}_2}
\newcommand{\E}{\mathbbm{1}}
\newcommand{\bk}{k}
\newcommand{\bG}{G}
\newcommand{\Tr}{\text{Tr}}
\newcommand{\tr}{\text{tr}}
\newcommand{\pf}{\text{Pf}}
\renewcommand{\d}{\mathrm{d}}
\renewcommand{\i}{\mathrm{i}}
\newcommand{\e}{\mathrm{e}}
\renewcommand{\l}{\left}
\renewcommand{\r}{\right}
\renewcommand{\H}{\mathcal{H}}
\newcommand{\Hsp}[1]{\H_{\rm sp #1}}
\newcommand{\F}{\mathcal{F}}
\newcommand{\C}{\mathcal{C}}
\newcommand{\bra}[1]{\left\langle #1\right|}
\newcommand{\ket}[1]{\left| #1\right\rangle}
\newcommand{\braket}[2]{\left\langle #1|#2\right\rangle}
\newcommand{\ketbra}[2]{\left|#1\right\rangle\left\langle #2\right|}
\newcommand{\gr}{\Psi}
\newcommand{\grket}{\ket{\gr}}
\newcommand{\grbra}{\bra{\gr}}
\newcommand{\R}{}
\newcommand{\pp}{\Pi_1}

\newcommand{\lso}{\lambda_{\rm SO}}
\newcommand{\lnu}{\lambda_{\nu}}
\newcommand{\lr}{\lambda_{\rm R}}
\newcommand{\sles}{SLES}
\newcommand{\pbc}{periodic boundary conditions}
\newcommand{\trim}{TRIM}
\newcommand{\bz}{BZ}
\renewcommand{\mathbb}{\mathbbm}

\newcommand{\fig}[1]{Fig.~\ref{#1}}
\newcommand{\Sec}[1]{Sec.~\ref{#1}}
\newcommand{\eq}[1]{Eq.~(\ref{#1})}

\graphicspath{{./}}

\begin{document}

\title{Relating the entanglement spectrum of noninteracting band insulators to their quantum geometry and topology 
}

\author{Markus Legner} 
\affiliation{
Institut f\"ur theoretische Physik, ETH Z\"urich,
8093 Z\"urich, Switzerland
            }  

\author{Titus Neupert} 
\affiliation{
Institut f\"ur theoretische Physik, ETH Z\"urich,
8093 Z\"urich, Switzerland
            }  
\affiliation{
Condensed Matter Theory Group, 
Paul Scherrer Institute, 5232 Villigen PSI,
Switzerland
            } 


\date{\today}

\begin{abstract}
We study the entanglement spectrum of noninteracting band insulators, which can be computed from the two-point correlation function, when restricted to one part of the system.
In particular, we analyze a type of partitioning of the system that maintains its full translational symmetry, by tracing over a subset of local degrees of freedom, such as sublattice sites or spin orientations. The corresponding single-particle entanglement spectrum is the band structure of an entanglement Hamiltonian in the Brillouin zone. 
We find that the hallmark of a nontrivial topological phase is a gapless entanglement spectrum with an ``entanglement Fermi surface."
Furthermore, we derive a relation between the entanglement spectrum and the quantum geometry of Bloch states contained in the Fubini-Study metric. The results are illustrated with lattice models of Chern insulators and $\zz$ topological insulators.
\end{abstract}

\maketitle

\section{Introduction}\label{s:intro}

Quantum entanglement can be used to classify gapped phases of matter beyond the paradigm of Ginzburg-Landau-type symmetry breaking. For example, the presence of long-range entanglement between different parts of a system is in one-to-one correspondence with the notion of its topological order.\cite{wen-2002,chen-2010}
Alternative characterizations of topological order, for example via the state's content of fractionalized quasiparticles and their statistics, show that there exist different inequivalent types of topological ordered states. One expects that these different types of topological order manifest themselves in inequivalent ``patterns'' of long-range entanglement.
The crucial question is thus: How can the ``pattern'' of long-range entanglement of a quantum state be detected and how are its universal features separated from the nonuniversal ones?
Generically, this question is addressed by studying the entanglement properties between the two parts 1 and 2 that follow from some bipartitioning of the system.~\cite{bennet-1996,vidal-2003} 
The bipartite entanglement entropy can be computed from the reduced density matrix $\rho_{\mathrm{1}}$, which is obtained by tracing out the degrees of freedom in part 2.
In a seminal paper~\cite{haldane-2008}, Li and Haldane proposed to consider the full spectrum of the reduced density matrix $\rho_{\mathrm{1}}$ to obtain much richer information about the bipartite entanglement of the system. For example, the entanglement spectrum along a spatial cut reveals the universal counting of low-energy excitations of the boundary conformal field theory of the state. \cite{chandran-2011,qi-2012}

The abovementioned notion of topological order renders all states without long-range entanglement topologically trivial. However, such short-range entangled states can still have topological attributes that are protected by some symmetry of the system, constituting so-called symmetry-protected topological (SPT) phases.\cite{gu_wen-2009}
A subset of the SPT states that is particularly well understood are the systems of noninteracting fermions with translational symmetry, i.e., band insulators. 
The information about their short-range entanglement is entirely contained in the manifold of Bloch states of occupied bands.
This manifold becomes a metric space upon introducing a quantum distance measure between 
Bloch states at different lattice momenta in the Brillouin zone.\cite{provost-1980,marzari-1997} 
The corresponding Hermitian and gauge invariant quantum metric tensor contains both universal and nonuniversal information about the system. For one, the Chern number as a topological quantum number is obtained by integrating the imaginary part of the quantum metric over the Brillouin zone.\cite{TKNN,simon-1983,berry-1984} On the other hand, the trace of the quantum metric has been proposed as a measure for the ``complexity'' of the wave function.\cite{marzari-1997}
Intuitively, it measures by how much the entanglement of local degrees of freedom in the Bloch states changes as a function of momentum.

This description of a band insulator suggests the existence of intimate connections between geometry, topology, and entanglement. The relation between its topology and the entanglement spectrum has been studied for spatial bipartitionings of SPT states in various publications.\cite {fidkowski-2010,turner-2010-prb,hughes-2011,huang-2012-prb,huang_2012b,fang-2013,hsieh-2013}
It has been shown for noninteracting systems that the so-called entanglement Hamiltonian $H^{\mathrm{e}}:=-\log\,\rho_{\mathrm{1}}+{\rm const}$ is a Hermitian single-particle operator, which is related to the two-point correlation function of the ground state \cite{peschel-2003}. This Hamiltonian then supports gapless protected boundary modes, if the system is topologically nontrivial.
In this paper, we extend the analysis to bipartitionings that preserve the full translational invariance of the system. 
Using the examples of a Chern insulator and a $\zz$ topological insulator in two-dimensional space, we find that whenever the system is a topologically nontrivial insulator, the entanglement Hamiltonian describes a metal with an ``entanglement Fermi sea.'' 
Furthermore, we draw a connection between the quantum geometry and the entanglement spectrum by showing that the trace over the entanglement Hamiltonian is equal to the quantum distance measure between Bloch states.   

In the remainder of this introduction, we are going review the entanglement spectrum and the quantum geometry of band insulators to set up the notation. In \Sec{s:results}, we will derive the announced results and illustrate them with concrete examples in \Sec{s:ex}.

\subsection{The entanglement spectrum}\label{s:es}
Consider a many-body quantum system with a unique, gapped ground state $\grket$ that is defined on a Fock space $\F$, built from a single-particle Hilbert space $\Hsp{}$ of dimension $\mathrm{dim}(\Hsp{})=\mathsf{N}$.
A partitioning of the system in two parts 1 and 2 is defined by a partitioning $\Hsp{}=\Hsp{1}\oplus\Hsp{2}$ inducing a decomposition $\F=\F_1\otimes\F_2$ of the Fock space. 
Then, the ground state $\grket$ can be written as a Schmidt decomposition
\begin{equation}
\grket=\sum_\alpha\frac{1}{\sqrt{Z}}\exp\l(-\frac{E_\alpha^\e}{2}\r)\ket{\alpha,1}\otimes\ket{\alpha,2}\ ,
\end{equation}
where $\ket{\alpha,i}\in\F_i$, $i=1,2$, and the factor $Z^{-1/2}$ with $Z:=\sum_\alpha\e^{-E_\alpha^\e}$ ensures the normalization. We can define the reduced density matrix of the first subsystem 
\begin{equation}
\rho\R_1:=\Tr_2\grket\grbra
\end{equation}
by performing the partial trace $\Tr_2$ over the degrees of freedom in $\F_{\mathrm{2}}$. The reduced density matrix can be represented as the exponential of a Hermitian operator,  the so-called entanglement Hamiltonian $H^\e_1$:
\begin{equation}
\rho\R_1=
\frac{1}{Z^\e_1}\e^{-H^\e_1}\ ,
\quad
Z^\e_1:=\Tr_1\, \e^{-H^\e_1}\ .
\end{equation}
The eigenvalues $E_\alpha^\e$ of this entanglement Hamiltonian are the entanglement spectrum. It has been shown that for an appropriate spatial cut, gapless edge modes of the energy spectrum lead to degenerate entanglement eigenvalues.\cite{fidkowski-2010}

\medskip

We now specialize to systems of noninteracting fermions with a unique ground state. Then the Hamiltonian is bilinear in the second quantized fermionic operators and $\grket$ is a single Slater determinant. 
In this case, the entanglement Hamiltonian $H^\e_1$ is a bilinear in the fermionic operators as well.\cite{peschel-2003}
If we denote with $c^\dagger_{\mathsf{n}},\ \mathsf{n}=1,\cdots, \mathsf{N}$, the creation operators for single-particle states in $\Hsp{}$ in some basis,
the Hamiltonian $H$ and the entanglement Hamiltonian $H^\e_1$ have the representations
\begin{equation}
H=\sum_{\mathsf{m},\mathsf{n}=1}^{\mathsf{N}}c^\dagger_{\mathsf{m}}h^{\ }_{\mathsf{m},\mathsf{n}}c^{\ }_{\mathsf{n}}\,,\qquad
H^\e_1=\sum_{\mathsf{m},\mathsf{n}=1}^{\mathsf{N}}
c^\dagger_{\mathsf{m}}h^{\e}_{1;\mathsf{m},\mathsf{n}}c^{\ }_{\mathsf{n}}\,.
\end{equation}
It has been shown\cite{peschel-2003} that the matrix elements $h^{\e}_{1;\mathsf{m},\mathsf{n}}$ of the entanglement Hamiltonian are related to the restricted correlation matrix $C\R_1=P_1CP_1$ via
\begin{equation}
C^{\R}_1=\frac{1}{1+\exp\l(h^{\e}_{1}\r)}\ .\label{eq:C and he}
\end{equation}
Here, $P_1$ denotes the projection on the first subsystem $\Hsp{1}$, represented as a $\mathsf{N}\times \mathsf{N}$ matrix in $\Hsp{}$, and the two-point correlation function $C$ is a $\mathsf{N}\times \mathsf{N}$ matrix with entries $C_{\mathsf{m},\mathsf{n}}=\grbra c^\dagger_{\mathsf{m}}c_{\mathsf{n}}\grket$.
The correlation matrix $C$ is nothing but the matrix representation of the projector on the single-particle states that are occupied in $\grket$ in the single-particle Hilbert space $\Hsp{}$. In systems with translational invariance, when the states can be labeled by some pseudomomentum $k$ (see below), we will call this projection $\Pi(k)$.

Therefore, if $\grket$ is a gapped ground state, we can define a topologically equivalent flatband version of the Hamiltonian $h$ by
\begin{equation}
Q:=\E/2-C\ ,
\end{equation}
which has eigenvalues $-1/2$ ($+1/2$) for the single-particle states that are occupied (empty) in $\grket$. The restriction of this matrix to subsystem~1 can be defined as
\begin{equation}
Q\R_1:=\E/2-C_1=
\E/2-P_1C P_1\ ,
\end{equation}
which has eigenvalues $\lambda_{\mathsf{m}}\in\l[-1/2,1/2\r],\ {\mathsf{m}}=1,\cdots, \mathsf{N}$.\cite{turner-2010-prb} 

According to \eq{eq:C and he}, the eigenvalues $\lambda_{\mathsf{m}}$ are monotonously related to the eigenvalues $\epsilon_{\mathsf{m}}$ of $h^\e_1$ via
\begin{equation}
\lambda_{\mathsf{m}}=\frac{1}{2}\tanh\l(\frac{\epsilon_{\mathsf{m}}}{2}\r),
\qquad \mathsf{m}=1,\cdots, \mathsf{N},
\end{equation}
and the entanglement spectrum $E^\e_\alpha$ can be found by
\begin{equation}
E_\alpha^\e=\sum_{\mathsf{m}=1}^{\mathsf{N}} n_\alpha^{(\mathsf{m})}\epsilon_{\mathsf{m}}\ ,\label{eq:multi-particle entanglement}
\end{equation}
with all possible combinations of occupation numbers $n_\alpha^{(\mathsf{m})}\in\{0,1\}$ of the $\mathsf{m}$-th single-particle state\cite{turner-2010-prb}. Due to this direct correspondence of $\lambda_{\mathsf{m}}$ and $E^\e_\alpha$, in this paper we will refer to the values $\lambda_{\mathsf{m}}$ as the single-particle entanglement spectrum.
Note that as $\tr(P_1)<\mathsf{N}$, the spectrum of $C_1$ will trivially contain the value $0$ multiple times, such that only $\tr(P_1)$ eigenvalues contain information. The vanishing eigenvalues do not contribute to the entanglement spectrum as they correspond to $\lambda=1/2$ and $\epsilon=\infty$. Throughout this paper they will be omitted by working in the subspace $\text{Im}(P_1)$.

Observe that every vanishing eigenvalue $\lambda_{\mathsf{k}}=0$ leads to a two-fold degeneracy of all entanglement eigenvalues, as it implies $\epsilon_{\mathsf{k}}=0$, and $E_\alpha^\e=E_{\alpha'}^\e$ is left unchanged by replacing $n_\alpha^{(\mathsf{k})}=0$ with $n_{\alpha'}^{(\mathsf{k})}=1$.

One strength of the entanglement spectrum lies in the fact that if the Hamiltonian $h$ and the projector $P_1$ share a symmetry $S$, i.e., $[h,S]=[P_1,S]=0$, the entanglement spectrum can be ordered by the eigenvalues of $S$, for $S$ is then also a symmetry of $h^{\e}_1$. 
Here, we will consider cases where $P_1$ preserves the translational symmetries of $h$, either along one or two directions of two-dimensional space, so that the lattice momentum $k$ along these directions can be used to label the eigenstates and the entanglement eigenvalues. In this case, the projector $P_1$ is block diagonal where we call its $N\times N$ blocks $\pp(k)$.
Now we can write the Hamiltonian as
\begin{equation}
H=\sum_k\sum_{\alpha,\beta=1}^Nc^\dagger_\alpha(k)h_{\alpha,\beta}(k)c_\beta(k)
\end{equation}
and a spectral decomposition of the Bloch matrix $\left(h_{\alpha,\beta}\right)(k)$ is given by
\begin{equation}
h(k)=\sum_{a=1}^N\ket{u_a(k)}\varepsilon_a(k)\bra{u_a(k)}\ ,
\end{equation}
revealing the Bloch states $\ket{u_a(k)}$, where the bands are labeled by the index $a=1,\dots,N$. $N$ is the total number of Bloch bands, and $k\in\mathrm{BZ}$ is a momentum in the first Brillouin zone (BZ). The projector on the $n<N$ occupied bands is then given by
\begin{equation}
\Pi(k):=\sum_{a=1}^n\ketbra{u_a(k)}{u_a(k)}\ .
\end{equation}
For $\tr(\pp)=m$ the spectrum of $\pp\Pi(k)\,\pp$ 
will contain the value $0$ at least $N-m$ times. As discussed above, we will omit these values, which do not contain any useful information, by just working in the subspace $\text{Im}(\pp)$. Note that we use $\tr$ for the trace of matrices to distinguish it from the trace $\Tr$ in the Fock space $\F$.

\subsection{Quantum geometry}\label{s:qg}
The single-particle eigenstates $\ket{u_a(k)}$ are elements of the complex projective space $\mathbb{CP}^{N-1}$. The Fubini-Study distance on $\mathbb{CP}^{N-1}$
can be used to define a distance measure  
\cite{provost-1980,pati-1991,marzari-1997}
\begin{equation}
d(k_1,k_2):=\sqrt{n-\sum_{a,b=1}^{n}\left|\braket{u_a(k_1)}{u_b(k_2)}\right|^2}
\label{eq: distance between states}
\end{equation}
between states at momenta ${k}_1$ and ${k}_2$ for the lowest $n$ bands occupied, the so-called quantum distance.
The expression~\eqref{eq: distance between states} can be reexpressed using $\Pi(k)$ as
\begin{equation}
\begin{split}
d^2(k_1,k_2)&=\tr\left\{\l[1-\Pi(k_1)\r]\Pi(k_2)\right\}\\
&=n-\tr\left[\Pi(k_1)\Pi(k_2)\right]\ .
\end{split}\label{eq:trace equal dim}
\end{equation}

An expansion of the infinitesimal distances
\begin{equation}
d^2(k,k+\d k)=\sum_{\mu\nu}g_{\mu\nu}(k)\,\d k^\mu\d k^\nu\ ,\label{eq:metric}
\end{equation}
defines the Riemannian metric $g_{\mu\nu}$. This tensor is the symmetric part of the Fubini-Study metric tensor (also called quantum geometric tensor)
\begin{equation}
q_{\mu\nu}=g_{\mu\nu}-\frac{\i}{2}f_{\mu\nu}\ ,
\end{equation}
with its antisymmetric part $f_{\mu\nu}$ being the Berry curvature.\cite{ma-2010}

\medskip

In general the distance between two projectors can be defined as
\begin{equation}
\begin{split}
d^2(P_1,P_2):&=\frac12\,\tr\l[\l(P_1-P_2\r)^2\r]\\
&=\frac12\l[\tr(P_1)+\tr(P_2)\r]-\tr(P_1P_2)\ ,
\end{split}
\end{equation}
which is induced by the Frobenius norm on the vector space of $N\times N$ matrices.
For projectors with equal traces $\tr P_1=\tr P_2=n$ this expression reduces to \eq{eq:trace equal dim}.
Note that for projectors with different traces it is not possible to define a metric tensor, for there exists no smooth interpolation between two projectors with unequal traces.

\section{Results}\label{s:results}
A commonly studied type of partitioning is a cut in position space, separating the system into a left and right half. Translational symmetry is only broken orthogonal to the cut, such that the pseudomomenta parallel to the cut remain good quantum numbers.
It has been shown that band insulators with nonzero Chern number
feature a gapless branch of chiral states in the entanglement spectrum, which are localized near the position-space cut. The single-particle entanglement eigenvalue of this branch of 
states flows from $\lambda=\mp1/2$ to $\lambda=\pm1/2$ as the momentum parallel to the cut is varied through the \bz.\cite{huang-2012-prb}

Here, we want to consider a different class of bipartitionings which leave the translational symmetries of the system unchanged.\footnote{A similar extensive cut has been considered in Ref.~\onlinecite{hsieh-2013}}  We will refer to the resulting entanglement spectrum as the sublattice entanglement spectrum (\sles). First, we will show how topological information is contained in the \sles. Thereafter we will relate the entanglement spectrum to the quantum distance between the ground state of our system and the projector of our bipartitioning, both for a spatial entanglement spectrum and the \sles.

\subsection{The sublattice entanglement spectrum}\label{s:part}
To define the \sles, one  introduces a partitioning of the internal degrees of freedom at every lattice site, such as sublattices, orbitals, or spin species, and traces out a subset of these on-site degrees of freedom.
As the \sles\ preserves the translational symmetry of the system, all components of $k$ remain good quantum numbers and we can write
\begin{equation}
Q_1\R(k)= \E/2-\pp\,\Pi(k)\, \pp\ .
\end{equation}
Here, all operators can be represented by $N\times N$ matrices, where the number of bands $N$ is equal to the number of internal degrees of freedom at each lattice site. 

We find that for a suitable choice of the bipartitioning of internal degrees of freedom, a nontrivial topological phase implies a gapless \sles\ covering (in the thermodynamic limit) all values in the interval $[-1/2,1/2]$. We observe that the lowest many-body entanglement eigenvalue $E^\e_{\rm min}$, according to \eq{eq:multi-particle entanglement}, is obtained by filling all single-particle states with $\lambda<0$. This suggests identifying these states with an ``entanglement Fermi sea'' and defining an ``entanglement Fermi surface'' at $\lambda=0$. A topological (Chern) insulator is therefore mapped to an ``entanglement metal'' (with an entanglement Fermi surface) and we are trading the topological stability provided by the band insulator's energy gap for the topological stability of this Fermi surface.\cite{volovik-2012}  We will exemplify this for two-dimensional systems of symmetry classes~A (Chern insulator) and AII ($\mathbbm{Z}_2$ topological insulator) in the notation of Altland and Zirnbauer\cite{altland-1997,schnyder-2008} in the following section.

In general the \sles\ can be studied for arbitrary projections on the internal degrees of freedom. However, not every choice of partitioning is useful in order to identify topological properties. For example one could think of two layers of a two-dimensional insulating system coupled weakly (as compared to their gaps) by interlayer hopping. The entanglement spectrum for a partitioning of these two layers will then only depend on the interlayer coupling and not on the topology of the two systems (see also Ref.~\onlinecite{hsieh-2013}). A similar example, using spin degree of freedom, will be presented for the Kane-Mele model in \Sec{ss:kmm}.

Also, we assumed the projection $\pp$ to be $k$-independent. This is a natural assumption emphasizing a clear physical interpretation of the bipartitioning but it is not strictly required. However, a necessary condition to a $k$-dependent projection is that it can be continuously deformed to a constant $\pp$; i.~e., it does not carry some nontrivial topology itself. An obvious choice for $\pp$ violating this condition would be the projection $\Pi(k)$ on the occupied bands. In this case we find $\lambda_i(k)=1/2$ everywhere, as $\left[\Pi(k)\right]^3=\Pi(k)$.

Despite these subtleties, the situation here is no worse than with the entanglement spectrum of a spatial bipartitioning. Also in this case there exist bipartitionings which do not contain topological information about the system. For example, a cut between Wannier states would not reveal the topology of the ground state, for the Wannier states are flowing with the entanglement eigenvalues.

\subsubsection{The \sles\ for Chern insulators}\label{ss:sles-chern}
First, we want to consider a two-dimensional model of class~A, where for simplicity we assume two bands, one of which is occupied. In this case we will prove that a nonvanishing Chern number implies a gapless \sles\ for \emph{any} choice of the bipartitioning. 

It has been shown that the Chern number of a Bloch band is given by the sum of the vorticities of all vortices of an arbitrarily chosen component $\alpha$ of the two-component vector $u(\bk)$ (note that we drop the band index $a$ as there is only one occupied band).\cite{hatsugai-1997,hatsugai-1999} The vorticity (or charge of a vortex) is there defined as the winding number of the phase of the $u_\alpha(\bk)$ around the vortex. Note that care must be taken to choose the gauge such that none of the components of $u(k)$ is multivalued at the vortices.

Consequently, within the topological phase with nonvanishing Chern number there must be at least one vortex at $\bk_\alpha^0$ for each component $\alpha$ of $u(\bk)$ in the \bz. For a two-band model ($\alpha=1,2$) the vectors $u(k_1^0)$ and $u(k_2^0)$ are accordingly orthogonal. By choosing 
\begin{equation}
\pp=\Pi(\bk_1^0)\label{eq:def projector chern}
\end{equation}
 as the projector on the occupied band at a vortex of the first component, we find that the single-particle entanglement eigenvalue obeys $\lambda(\bk_1^0)=-1/2$ and $\lambda(\bk_2^0)=+1/2$, proving our statement.

In fact any other nontrivial ($\pp\neq0$ and $\pp\neq\E$) choice of the bipartitioning leads to the same result for the two-band model: Any nontrivial projector $\pp'$ is related to $\pp$ by a unitary transformation
\begin{equation}
\pp'=U\pp U^\dagger\ .
\end{equation}
The components of the vectors $Uu(\bk)$ have vortices at different points in the \bz, where the \sles\ with the new projector will have the values $\lambda=\pm 1/2$.

The reverse statement (a gapless \sles\ implies topological order) is in general not true: For example, it is possible to have two vortices with opposite vorticities adding up to a Chern number of zero. Nevertheless in this case we would find a gapless \sles.

This argument can in principle be generalized to multiple bands. However, in this case we will find that the topological information within the \sles\ depends on the chosen bipartitioning as argued above.

\begin{figure}[t]
\includegraphics[width=.46\textwidth]{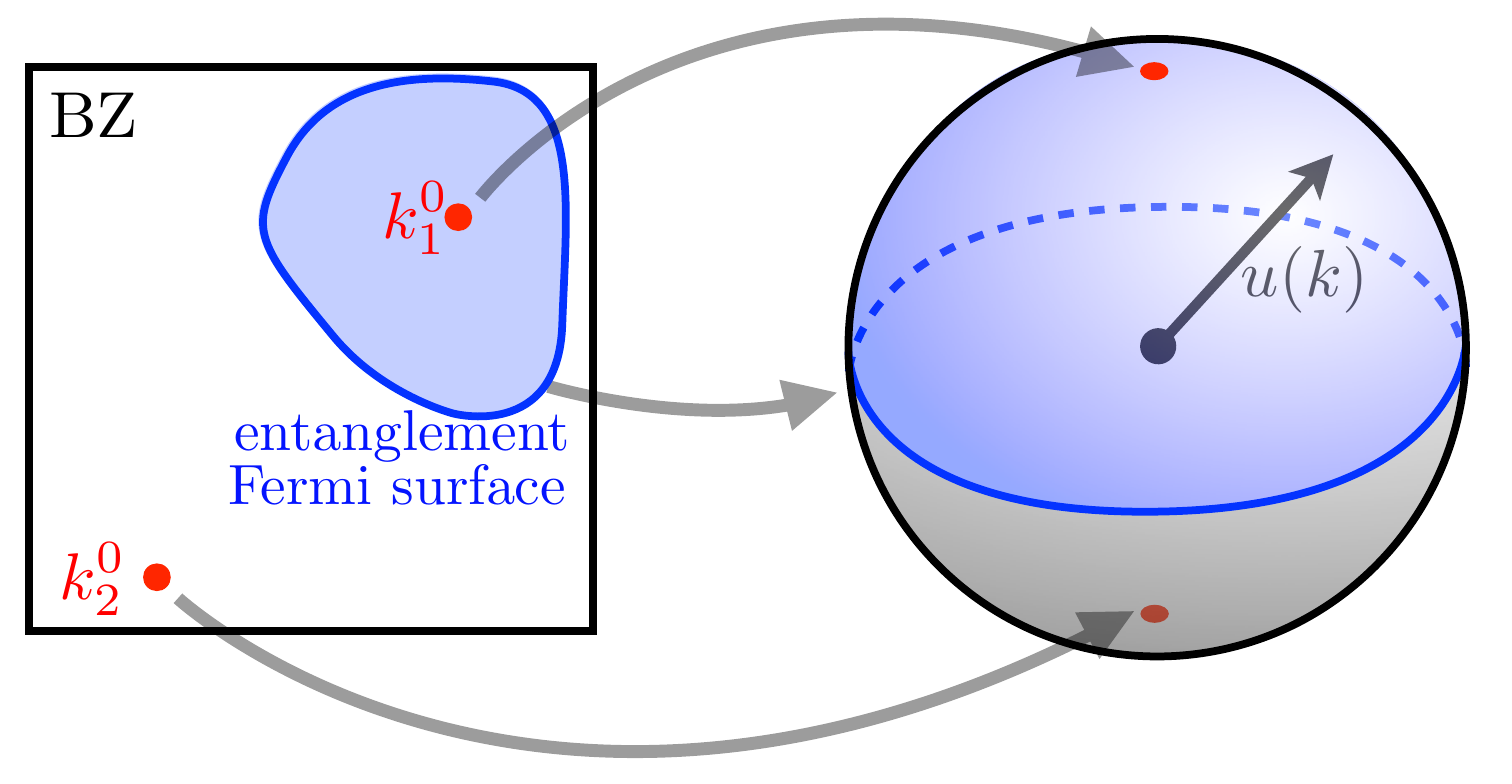}
\caption{(Color online) Mapping from the \bz\ to the Bloch sphere for 
the bipartitioning defined by the projector $\pp$ [\eq{eq:def projector chern}] in the case of a band with nonvanishing Chern number.
The points $k_1^0$ and $k_2^0$, where first and the second component of the Bloch spinor vanish, are mapped to the north and south pole of the Bloch sphere, respectively. The entanglement Fermi surface is mapped to the equator. Any other choice of $\pp$ corresponds to a rotation of the Bloch sphere, such that the mapping from the \bz\ to the Bloch sphere changes.
However, the existence of the entanglement Fermi surface is guaranteed by the nonvanishing Chern number for any choice of $\pp$.
}
\label{fig:bloch}
\end{figure}

\medskip

For a two-band model, the relation between a nonvanishing Chern number and the existence of an entanglement Fermi surface can be visualized using the Bloch sphere, parametrized by the polar and azimuthal angles $\theta$ and $\phi$, respectively (see \fig{fig:bloch}). The state vector $u(k)$ of the occupied band can be represented by a point on the unit sphere $S^2$. We define a particular mapping from the \bz\ to the unit sphere in such a way that the image of the projector $\pp$ is represented by the north pole $\theta=0$ of the sphere. This can be accomplished by writing the state vector as
\begin{equation}
u(k)=
U^{\ }_{\pp}
\begin{pmatrix}\cos\l[{\theta(k)}/{2}\r]\e^{\i{\phi(k)}/{2}}\\\sin\l[{\theta(k)}/{2}\r]\e^{-\i{\phi(k)}/{2}}
\end{pmatrix}\ ,
\end{equation}
where $U_{\pp}$ is a $2\times2$ unitary matrix that diagonalizes $\pp$, that is,
$U^{\dagger}_{\pp} \pp^{\ } U^{\ }_{\pp}=\mathrm{diag}(1,0)$.
Then the entanglement eigenvalue is given by
\begin{align}
\lambda(k)=1/2-\cos^2\l[\theta(k)/2\r]=-1/2\cos\l[\theta(k)\r]\ ,
\end{align}
which takes the value $-1/2$, $1/2$, and $0$ at the north pole, south pole, and equator, respectively. Thus, the entanglement Fermi surface maps to the equator of the Bloch sphere.  

In two dimensions, the Chern number can be regarded as the winding number of the mapping $u(k)$ from the \bz\ to the unit sphere. Thus a nonzero Chern number implies that for every point $v$ on the unit sphere there is a $k_v$ in the \bz\ such that $u(k_v)=v$. In particular this implies that for every $\lambda\in[-1/2,1/2]$ there exists a point in the \bz\ such that $\lambda(k)=\lambda$. 

Note that this argument applies to systems with arbitrary Chern numbers $\C$. For $|\C|>1$, the Bloch vector will wind around the sphere multiple times, leading to multiple points mapped to the north pole, south pole, and equator. Therefore, the number of entanglement Fermi surfaces is equal to or larger than $|\C|$ in this case.

Again, we find that a vanishing Chern number does not necessarily lead to a gapped \sles: The \sles\ is gapless if and only if the mapping $\theta(k)$ from the \bz\ to the Bloch sphere is surjective. However, surjectivity does not imply a nontrivial winding, such that a gapless \sles\ can be present even for a vanishing Chern number.

Note that the original choice~(\ref{eq:def projector chern}) of $\pp$ is already diagonal, i.~e., $U_{\pp}=\E$. Choosing a different bipartitioning rotates the Bloch sphere in this picture, leading to different mappings $\theta(k)$ and $\phi(k)$. However, due to the winding of the Bloch vector for a nonzero Chern number, the \sles\ is gapless for any choice of $\pp$.

\subsubsection{The \sles\ for $\zz$ topological insulators}\label{ss:sles-topological}

Now we want to analyze topological information in the \sles\ of a two-dimensional system of class AII. This symmetry class is characterized by time-reversal symmetry
\begin{equation}
\Theta h(k)\Theta^{-1}=h(-k)\ ,
\end{equation}
where the operator $\Theta$ is antiunitary and fulfills $\Theta^2=-1$ (which implies $\Theta^\dagger=\Theta=-\Theta^{-1}$). This property also leads to a Kramer's degeneracy of the energy bands at the four time-reversal invariant momenta (\trim) $k=\Gamma_i,\ i=1,\dots, 4$,
where $-\Gamma_i=\Gamma_i+\bG$ for a reciprocal lattice vector $\bG$.

The $\zz$ invariant can be expressed in terms of zeros of the Pfaffian~\cite{kane-mele-2005a}
\begin{equation}
p(\bk):=\pf\left[m(k)\right]\ ,
\end{equation}
where the antisymmetric matrix $m(k)$ is defined as
\begin{equation}
m_{a,b}(k):=\bra{u_a(\bk)}\Theta\ket{u_b(\bk)}\ ,
\end{equation}
with $a$ and $b$ labeling occupied bands.
If there is a discrete set of $n_0$ points with $p(\bk)=0$, the $\zz$ invariant is given by
\begin{equation}
\nu=\frac{n_0}{2}\quad\text{mod } 2\ ,
\end{equation}
while in general it can be expressed as the winding number of the phase of $p(\bk)$ along a time-reversal invariant path $\gamma$ enclosing \emph{half} the \bz.\cite{kane-mele-2005a}

The minimal example for such a $\zz$ topological insulator has four bands two of which are occupied.\footnote{Kramer's theorem implies a gapless energy spectrum for a two-band model.} The Pfaffian in this case is just given by
\begin{equation}
p(\bk)=m_{12}(k)=\bra{u_1(\bk)}\Theta\ket{u_2(\bk)}\ .
\end{equation}
At $\bk^*$ where $p(\bk^*)=0$ we find that $\ket{u_2(\bk^*)}$ and $\Theta\ket{u_1(\bk^*)}$ are orthogonal. Further, $\ket{u_1(\bk^*)}$ and $\Theta\ket{u_1(\bk^*)}$ are orthogonal due to $\Theta^2=-1$. [Any antiunitary operator $U$ fulfills $\braket{Uu}{Uv}=\braket{v}{u}$. For $v=Uu$ and $U^2=-1$ we find $\braket{Uu}{Uv}=\braket{Uu}{U^2u}=-\braket{Uu}{u}$ and $\braket{v}{u}=\braket{Uu}{u}$. This leads to $\braket{Uu}{u}=-\braket{Uu}{u}$ implying $\braket{Uu}{u}=0$.] Therefore we can define the bipartitioning as the projection 
\begin{equation}
\pp=\ket{u_1(\bk^*)}\bra{u_1(\bk^*)}+\Theta\ket{u_1(\bk^*)}\bra{u_1(\bk^*)}\Theta^{-1}\ .\label{eq:projection}
\end{equation}
Then, at this point $\bk^*$ we find
\begin{equation}
\Pi(\bk^*)\pp\Pi(\bk^*)=\ket{u_1(\bk^*)}\bra{u_1(\bk^*)}\ ,
\end{equation}
as $\pp\ket{u_1(\bk^*)}=\ket{u_1(\bk^*)}$ and $\pp\ket{u_2(\bk^*)}=0$. For arbitrary projectors $P_1$ and $P_2$ the spectrum of $P_1P_2P_1$ is equal to that of $P_2P_1P_2$.\footnote{$P_1P_2P_1 u=\lambda u$ for $\lambda\neq0$ implies $u\in\text{Im}(P_1)$ and therefore $P_1P_2u=\lambda u$ as well as $P_2u\neq0$. Multiplying with $P_2$ and using $P_2^2=P_2$ we find $P_2P_1P_2(P_2u)=\lambda(P_2u)$.} Therefore the entanglement spectrum at $\bk^*$ is given by 
\begin{equation}
\lambda_1(\bk^*)=-\lambda_2(\bk^*)=1/2\ .\label{eq:es-dirac}
\end{equation} 

We note that our definition of $\pp$ is time-reversal invariant due to 
\begin{align}
\Theta \pp\Theta^{-1}&=\Theta\ket{u_1(\bk^*)}\bra{u_1(\bk^*)}\Theta^{-1}+\ket{u_1(\bk^*)}\bra{u_1(\bk^*)}\nonumber\\
&=\pp\ .
\end{align}
Thus Kramer's theorem implies a degeneracy of the eigenvalues of $\pp\,\Pi(\Gamma_i)\,\pp$ at the \trim, leading to 
\begin{equation}
\lambda_1(\Gamma_i)=\lambda_2(\Gamma_i)\ .\label{eq:es-trim}
\end{equation}
Equations~(\ref{eq:es-dirac}) and (\ref{eq:es-trim}) together with the continuity of the $\lambda_i$ as a function of $k$ then imply a gapless \sles\ for the bipartitioning~(\ref{eq:projection}), if the system is a $\zz$ topological insulator. 

Again the converse is  \emph{not} true in general, as there could be multiple pairs of points with $p(\bk)=0$. There are other choices of $\pp$ leading to the same topological information in the \sles\ as we will see in the example of the Kane-Mele model in \Sec{s:ex}.

\subsection{Entanglement spectrum and quantum geometry}\label{s:es-qg}
In \Sec{s:qg} we have defined the quantum distance for any pair of projectors with in general different traces $\tr (P_1)\neq \tr (P_2)$. We can now investigate specifically the distance between the projection on the occupied bands $\Pi(k)$ and the projection $\pp$ of the bipartitioning in the entanglement spectrum. We will assume that we have $\tr\l[\Pi(k)\r]=n$ occupied bands and that $\tr(\pp)=m$. The distance is then given by
\begin{equation}
d_{\pp}^2(k):=d^2(\pp,\Pi(k))=\tfrac{m+n}{2}-\tr\l[\pp\,\Pi(k)\r]\ .
\end{equation}
As $\pp^2=\pp$ and the trace is invariant under cyclic permutations, we can write
\begin{equation}
d_{\pp}^2(k)=\tfrac{m+n}{2}-\tr\l[\pp\,\Pi(k)\,\pp\r]\ .
\end{equation}
The trace of an operator is just the sum of all its eigenvalues which in this case are given by the entanglement spectrum. Therefore the distance is
\begin{equation}
\begin{split}
d_{\pp}^2(k)&=\tfrac{m+n}{2}-\sum_{i=1}^m\left(\frac12-\lambda_i(k)\right)\\
&=\frac{n}2+\sum_{i=1}^m\lambda_i(k)\ .
\end{split}
\label{eq: quantum distance}
\end{equation}
Note that we only sum the $m$ eigenvalues which do not generally vanish, as discussed at the end of \Sec{s:es}.

The quantum distance in this case measures the overlap of the ground state wavefunction with the chosen partitioning (where $d=0$ for full overlap). The single-particle entanglement spectrum is related to the overlap of one single-particle state with the partitioning ($\lambda=-1/2$ for full overlap). It is thus natural that the quantum distance between the occupied bands and any bipartitioning of a system is defined by the sum of the entanglement eigenvalues. 

\medskip

We can now use the (inverse) triangle inequality to derive a lower bound for the metric tensor in the \bz\ from the entanglement spectrum. The inequality reads for arbitrary projectors
\begin{equation}
d(k_1,k_2)\geq\left|d_{\pp}(k_1)-d_{\pp}(k_2)\right|\label{eq:dist-est}\ ,
\end{equation}
where we use the notation from above and have defined $d(k_1,k_2):=d(\Pi(k_1),\Pi(k_2))$.
Considering infinitesimal distances and the square of this inequality, we find
\begin{subequations}
\begin{equation}
\begin{split}
g_{\mu\nu}(k)\,\d k^\mu\d k^\nu&=d^2(k,k+\d k)\\
&\geq \left[d_{\pp}(k)-d_{\pp}(k+\d k)\right]^2\ ,\label{eq:div-dist}
\end{split}
\end{equation}
where the last expression can be expanded to
\begin{equation}
\begin{split}
&\textstyle\left(d_{\pp}(k)-\left\{d_{\pp}(k)^2-\sum\limits_{i=1}^m\left[\lambda_i(k+\d k)-\lambda_i(k)\right]\right\}^{1/2}\right)^2\\
&\textstyle=\left[d_{\pp}(k)-\left(d_{\pp}(k)^2-\sum\limits_{i=1}^m\partial_\mu\lambda_i(k)\d k^\mu\right)^{1/2}\right]^2\ .
\end{split}
\end{equation}
\end{subequations}

Assuming $d_{\pp}(k)\neq 0$, this leads to the estimate
\begin{subequations}
\begin{equation}
g_{\mu\nu}(k)\,\d k^\mu\d k^\nu\geq g_{\mu\nu}^{\pp}(k)\,\d k^\mu\d k^\nu\ ,\label{eq:tensor inequality}
\end{equation}
for arbitrary $\d k^\mu$, $\d k^\nu$. Here we defined the lower bound 
\begin{equation}
g_{\mu\nu}^{\pp}(k):=\frac{1}{4 d_{\pp}^2(k)}\sum_{i,j=1}^m\l[\partial_\mu\lambda_i(k)\r]\,\l[\partial_\nu\lambda_j(k)\r]\ ,\label{eq:geometry estimate}
\end{equation}
\end{subequations}
which depends on the projector $\pp$ representing the bipartitioning of the entanglement spectrum. 
In particular, the independence of the different components $\d k^\mu$ implies a lower bound for the diagonal elements of the metric tensor,
\begin{equation}
g_{\mu\mu}(k)\geq g_{\mu\mu}^{\pp}(k)\ .\label{eq:tensor diagonal inequality}
\end{equation}

For one specific bipartitioning the estimate~(\ref{eq:geometry estimate}) will generally not provide a good approximation to the quantum metric tensor at every point in the \bz. However, the inequalities~(\ref{eq:tensor inequality}) and (\ref{eq:tensor diagonal inequality}) hold for an \emph{arbitrary} projection $\pp$. 

Rewiring the expression~(\ref{eq:geometry estimate}) in terms of the distance $d_{\pp}(k)$ hides the relation with the entanglement spectrum but produces the simple form
\begin{equation}
g_{\mu\nu}^{\pp}(k)= d_{\pp}^2(k)\big\{\partial_\mu\log\l[d_{\pp}(k)\right]\big\}\big\{\partial_\nu\log\l[d_{\pp}(k)\right]\big\}\ .
\end{equation}

\medskip

The relation between quantum geometry and the entanglement spectrum is especially direct for the \sles\ when we choose a projection on one internal degree of freedom. In this case the entanglement spectrum is a single real number at every $k$ and we find
\begin{equation}
d_{\pp}^2(k)=\frac{n}2+\lambda(k)\ .\label{eq:quantum distance sles}
\end{equation}
If we consider a two-band model with $n=1$, this distance takes all values between $0$ and $1$ for a gapless entanglement spectrum.

In the case of the spatial entanglement spectrum (with open boundary conditions) there are discontinuities of the eigenvalues in a topological phase; see \Sec{ss:pfm} and Ref.~\onlinecite{huang_2012b}. These imply similar discontinuities in the quantum distance~\eqref{eq: quantum distance}. Using the estimate~(\ref{eq:dist-est}),
we find that also the quantum distance of arbitrarily close points in the \bz\ is finite. According to \eq{eq:metric} the metric tensor is singular at those points, reflecting the fact that no smooth gauge can be chosen in bands with nonzero Chern number. 

\section{Examples}\label{s:ex}
\subsection{The $\pi$-flux model}\label{ss:pfm}
The $\pi$-flux model is a 2-dimensional example of a Chern insulator.
\cite{haldane-1988,PFM-1989} It is defined on two interpenetrating square lattices and is characterized by a flux of plus/minus half the flux quantum through each plaquette (half the unit cell). This leads (in our chosen gauge) to imaginary NN hopping amplitudes of $t_1\e^{\pm \i\pi/4}$ (with $t_1\neq 0$). In addition, one introduces a NNN hopping amplitude of $\pm t_2$ and a staggered chemical potential of $\pm\eta$ on the two sublattices. 

For \pbc\ the Hamiltonian can be written as
\begin{equation}
h(\bk)=\boldsymbol{b}(\bk)\cdot\boldsymbol{\sigma}\ ,
\end{equation}
with the Pauli matrices $\boldsymbol{\sigma}=(\sigma_1,\,\sigma_2,\,\sigma_3)^t$ and
\begin{subequations}
\begin{align}
\begin{split}
b_1(k)+\i b_2(k)&=t_1\Big[\e^{-\i\frac{\pi}{4}}\l(1+\e^{\i(k_y-k_x)}\r)\\
&\hspace{10mm} +\e^{\i\frac{\pi}{4}}\l(\e^{-\i k_x}+\e^{\i k_y}\r)\Big]
\end{split}\\
b_3(k)&=2t_2\l(\cos k_x-\cos k_y\r)+\eta\ .
\end{align}
\end{subequations}
The energy spectrum of this model is gapless for $|\eta|=4|t_2|$. The system shows a topologically nontrivial phase for $\eta<4t_2$ with Chern number $\C=\text{sign}(t_1t_2)$.

\begin{figure}[t]
\centering
\includegraphics[width=.4\textwidth]{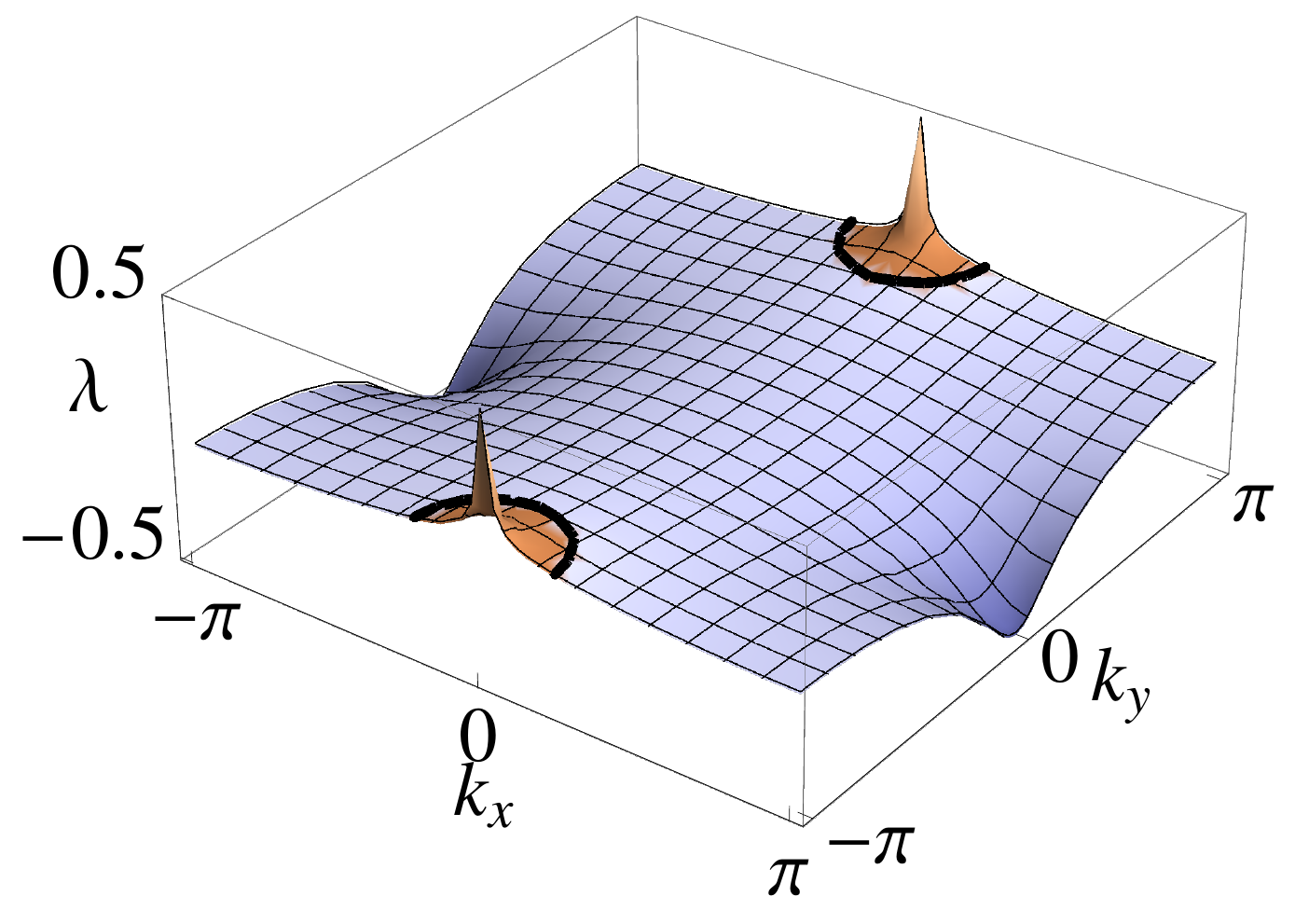}
\includegraphics[width=.4\textwidth]{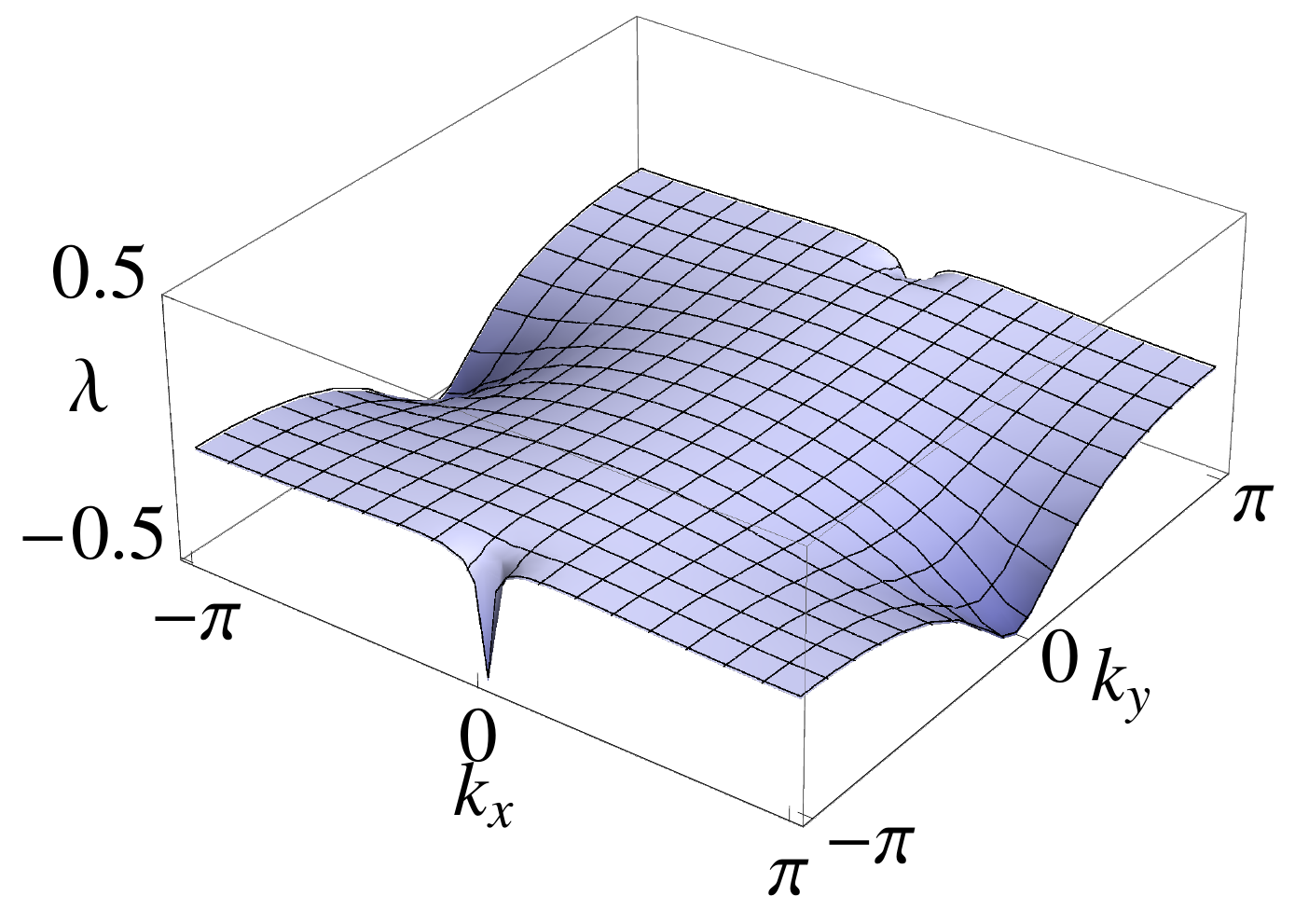}
\caption{(Color online) Plots of the \sles\ of the $\pi$-flux model for $t_2=0.1$ and $\eta=-0.35$ (top; topological), and $\eta=-0.45$ (bottom; trivial). The entanglement Fermi sea ($\lambda<0$) is shown in purple/light gray color, regions with $\lambda>0$ are shown in orange/dark gray, the thick black line shows the entanglement Fermi surface.
The entanglement spectrum is gapless (gapped) in the topological (trivial) case, with a phase transition at $\eta=-4t_2$.
According to \eq{eq:quantum distance sles} this plot also shows the quantum distance, as $\lambda(k)=d_{\pp}(k)-1/2$.}
\label{fig:pfm-sl}
\end{figure}

\subsubsection{The SLES for the $\pi$-flux model}
We now want to calculate the \sles\ for this model using \pbc. A natural choice of the bipartitioning is to trace over one sublattice, say sublattice B.\footnote{As discussed in \Sec{ss:sles-chern} any other choice would lead to the same qualitative results for this two-band model.}
The related projector $\pp$ on sublattice A reads as
\begin{equation}
\pp=\begin{pmatrix}
  1&0\\0&0
  \end{pmatrix}=\frac12\left(\sigma_0+\sigma_3\right)\ .
\end{equation}
The eigenstates are given by two-component vectors $u_a(\bk)=\left(u_a^{\rm A}(\bk),u_a^{\rm B}(\bk)\right)^t$ and we find a single entanglement eigenvalue which is given by
\begin{equation}
\lambda(\bk)=\frac{1}{2}-\left|u_1^{\rm A}(\bk)\right|^2\ .
\end{equation}
This is equal to $\lambda=\pm 1/2$ at the vortices of the two components (the poles of the Bloch sphere) and vanishes at the equator of the Bloch sphere. Figure~\ref{fig:pfm-sl} shows the \sles\ for different choices of $\eta$. As argued above, the \sles\ is gapless in the topological phases ($|\eta|<4|t_2|$) and becomes gapped in the topologically trivial phase. This is due to the change of the occupied state $u_1(\bk)$ at the point $\bk=(\pi,0)$ where the energy gap closes for $\eta=-4t_2$.

We note that for the $\pi$-flux model (being a two-band model), the quantum distance $d(\pp,\Pi(\bk))$ is given by \eq{eq:quantum distance sles}; it is just the \sles\ (being in this case a single real number) shifted by $+1/2$ and therefore contains exactly the same information as the \sles.

\begin{figure}[t]
\centering
\includegraphics[width=.4\textwidth]{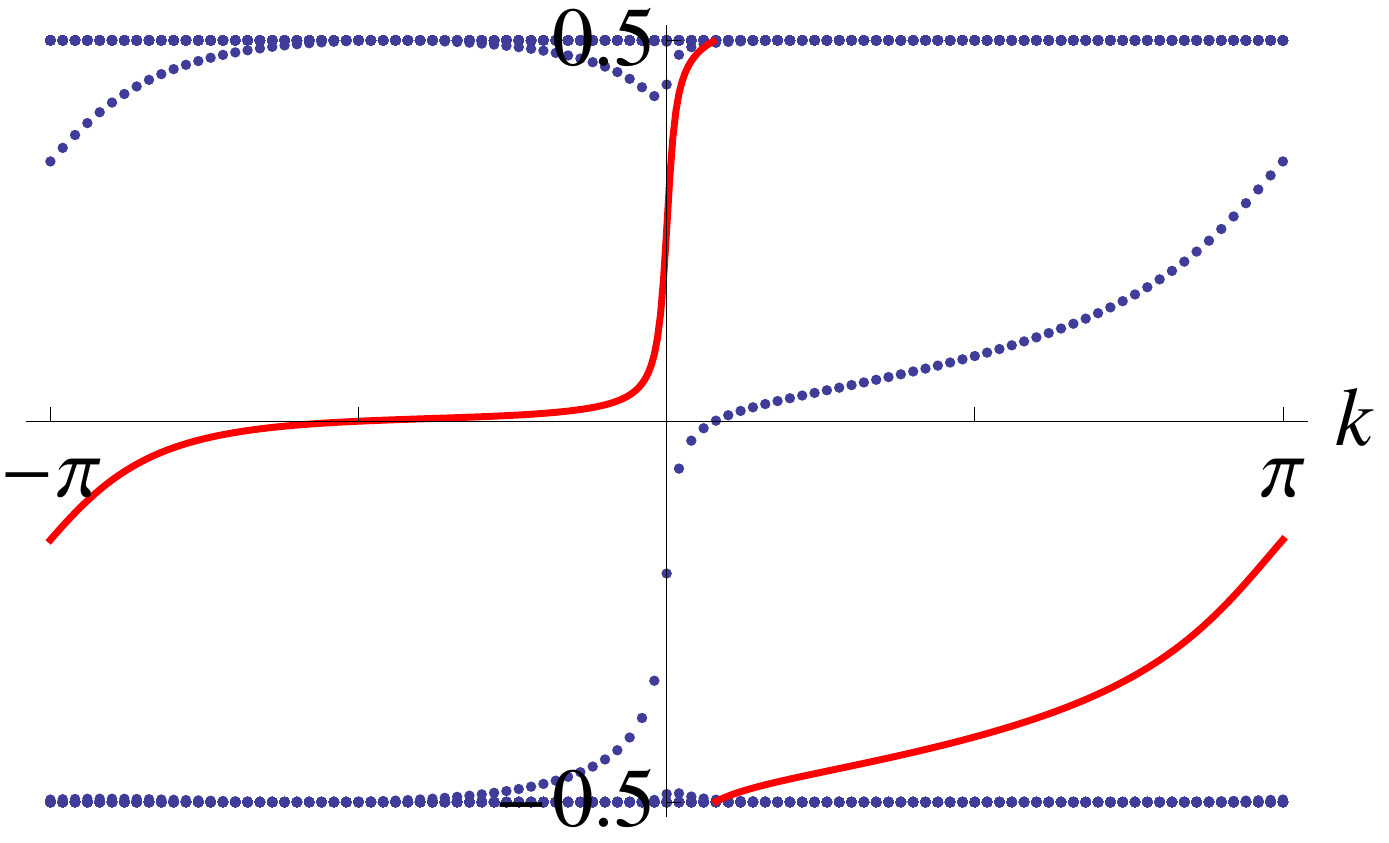}\\[2mm]
\includegraphics[width=.4\textwidth]{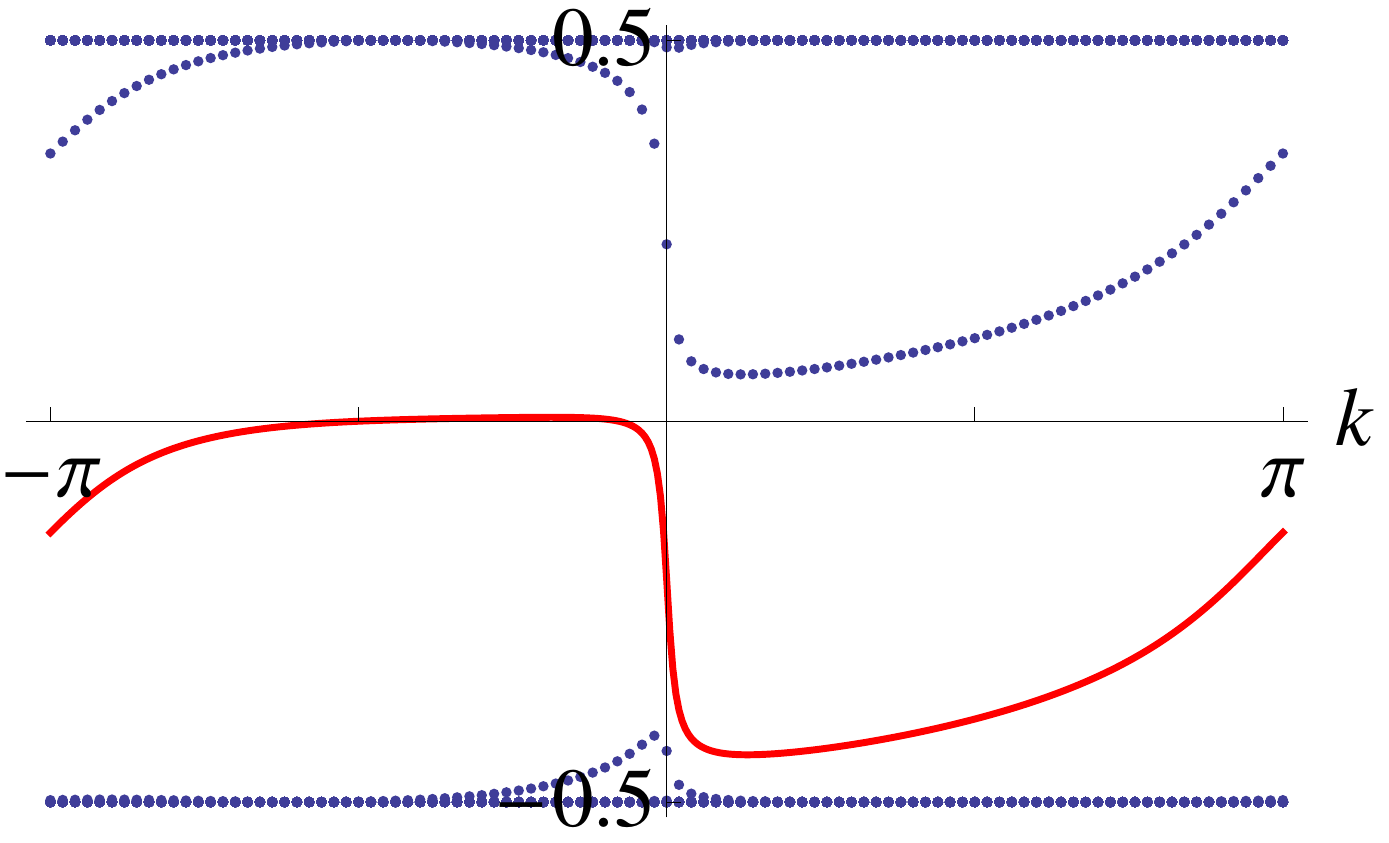}\\
\includegraphics[width=.3\textwidth]{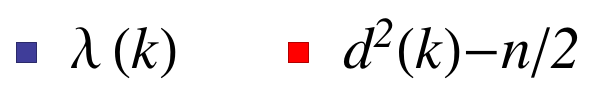}
\caption{(Color online) 
Single-particle entanglement spectrum $\lambda(k)$ (dots) and quantum distance $d^2(k)-n/2$ (continuous line)
as a function of the wavevector parallel to the applied spatial cut for the $\pi$-flux model. The cut is applied in the middle of a strip of $100$ lattice sites, i.~e., $N=200$ and $n=m=100$.
The parameters are the same as in \fig{fig:pfm-sl}, i.~e., $t_2=0.1$ and $\eta=-0.35$ (top), and $\eta=-0.45$ (bottom). The quantum distance is calculated using \eq{eq: quantum distance}.}
\label{fig:pfm}
\end{figure}

\subsubsection{The spatial entanglement spectrum for the $\pi$-flux model}

In addition to the SLES we can also calculate the single-particle entanglement spectrum and the quantum distance of a spatial bipartitioning, which is shown in \fig{fig:pfm}.
The band crossing in the entanglement spectrum as well as the discontinuity of the quantum distance are clearly visible in the topological phase and vanish in the trivial phase. We also observe that the discontinuous jump of one entanglement eigenvalue occurs precisely at the point where the crossing entanglement eigenvalue $\lambda=0$.
The spectral flow in the entanglement spectrum can be understood as a Wannier function moving across the cut introduced by the bipartitioning.\cite{huang-2012-prb} However, as $\lambda_i(2\pi)=\lambda_i(0)$, this spectral flow results in a discontinuity of at least one eigenvalue $\lambda$ which jumps from $\pm 1/2$ to $\mp 1/2$ and is connected to the jump of one Wannier state from one to the opposite edge.

In \fig{fig:pfm} we have chosen open boundary conditions orthogonal to the cut:
If \pbc\ are also imposed orthogonal to the cut, the bipartitioning results in two disconnected cuts in position space. This implies a spectral symmetry in the entanglement spectrum, with all single-particle entanglement eigenvalues coming in pairs $\pm\lambda$.\cite{huang-2012-prb} The use of open boundary conditions avoids this doubling of entanglement eigenvalues. Note that for a spatial bipartitioning with periodic boundary conditions along both directions the symmetry of the single-particle entanglement spectrum implies that the quantum distance is trivially $d_{\pp}^2(k)=n/2=\rm const$.

\subsection{The Kane-Mele model}\label{ss:kmm}
In order to study the quantum spin Hall effect in graphene, Kane and Mele introduced a tight-binding model on the honeycomb lattice\cite{kane-mele-2005a,kane-mele-2005b}. It consists of NN hopping $t$, spin-dependent NNN hopping (spin-orbit interaction $\lso$), a Rashba coupling $\lr$, and a staggered sublattice potential $\lnu$, which breaks the sublattice symmetry.

If we define the coordinates where $k_{1,2}:=1/2(k_x\pm\sqrt 3 k_y)$, the two Dirac points are at $\pm\bk^*=\pm(k^*_1,k^*_2)=\pm(2\pi/3,2\pi/3)$. The energy spectrum at the Dirac points is
\begin{align}\label{eq:kmm-ene}
\varepsilon_{i,j}(\pm\bk^*)=(-1)^i\,3\sqrt3\lso+(-1)^j\,\sqrt{\lnu^2+(9/4)\lr^2}
\end{align}
for $i,j=1,2$. Note that we replaced the band index $a=1,\dots,4$ by the two indices $i$ and $j$.

\begin{figure}[t]
\includegraphics[trim=0mm 0mm 0mm 1mm,clip=true,height=.22\textwidth]{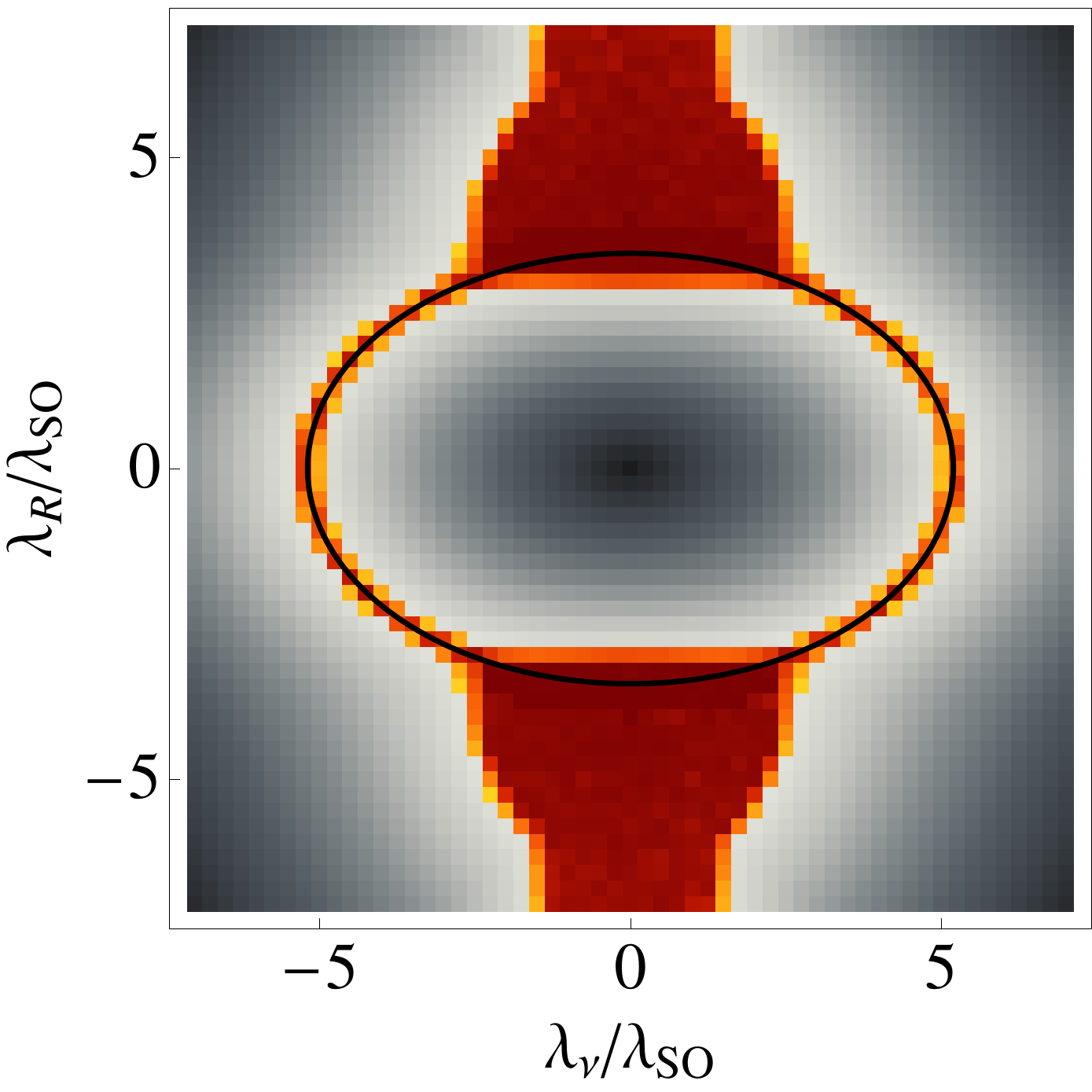}
\hspace{0mm}\includegraphics[trim=21mm 0mm 0mm 0mm,clip=true,height=.22\textwidth]{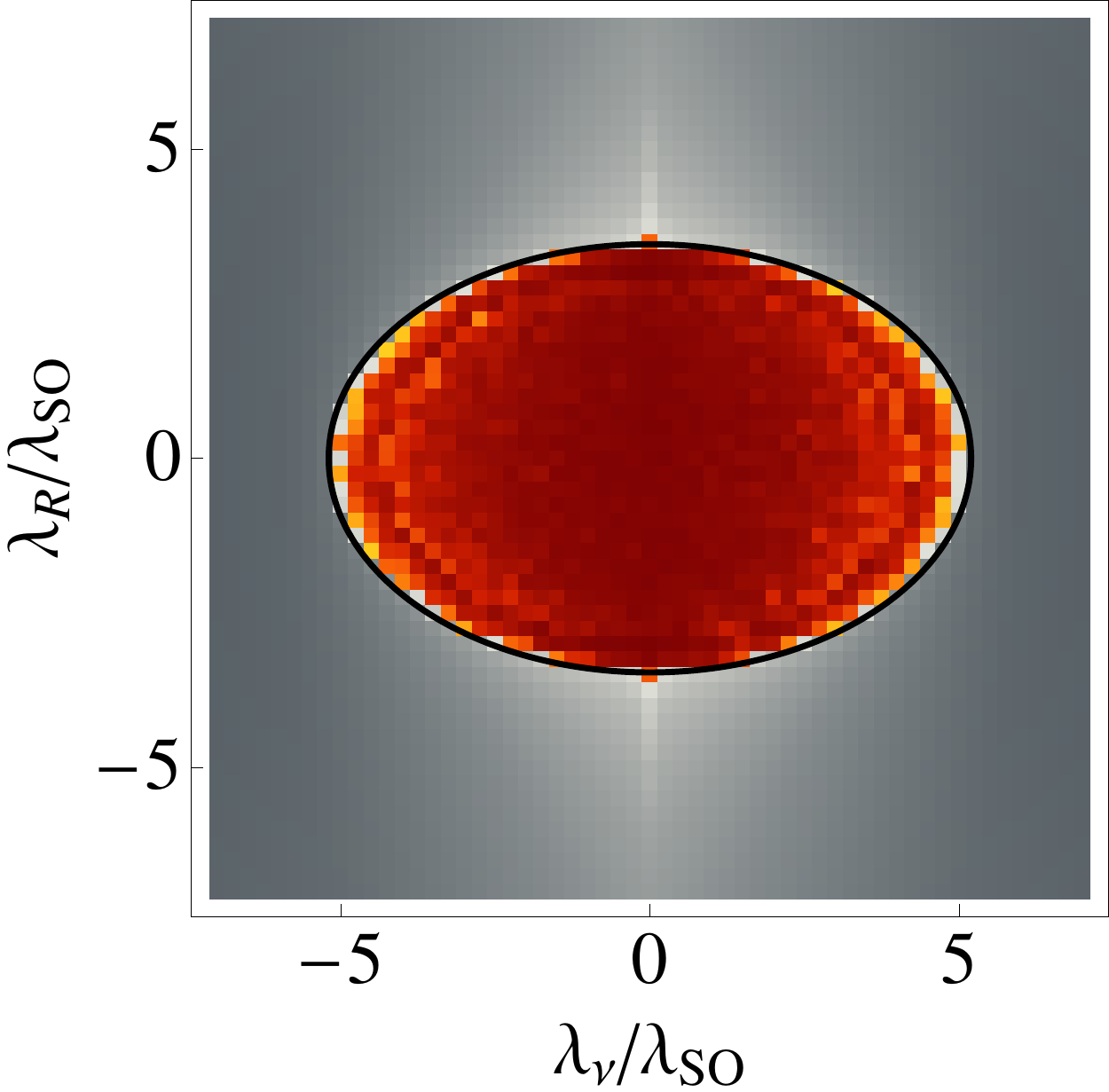}
\hspace{0mm}\raisebox{6mm}{\includegraphics[height=.18\textwidth]{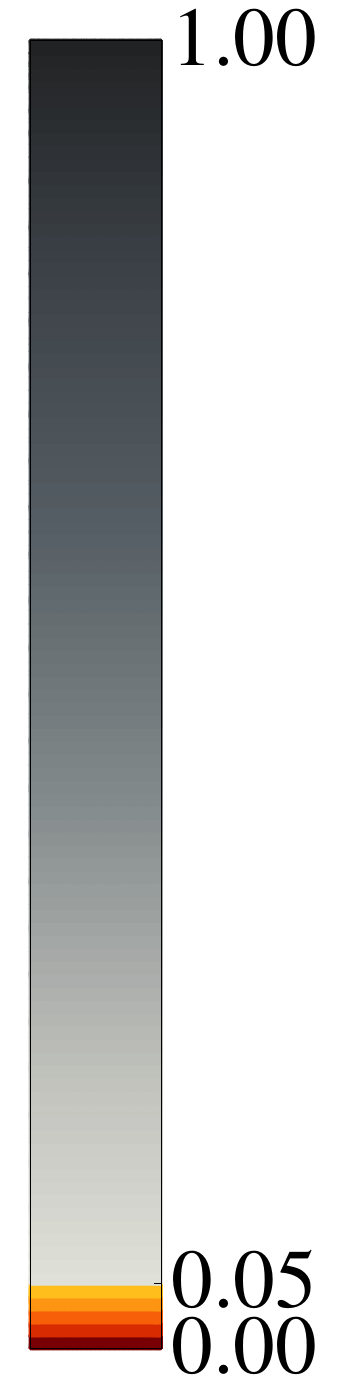}}
\caption{(Color online) Regions of gapless energy spectrum (left) and gapless \sles\  (right) in the Kane-Mele model for different values of the parameters $\lnu$ and $\lr$.
The plots were obtained by numerically evaluating the (entanglement) spectrum at random points in the first \bz\ for each set of parameters. The color coding shows the maximal distance between eigenvalues; red and yellow colors show values below $0.05$. Additional precision with a larger sample of points was used in a range around the phase transition.
The black contour shows the ellipse of \eq{eq:ellipse}.}
\label{fig:heatmap}
\end{figure}

The energy gap closes at $k^*$ when the parameters $\lnu$ and $\lr$ satisfy
\begin{subequations}
\begin{equation}
1=\frac{1}{27}\,\left(\frac{\lnu}{\lso}\right)^2+\frac{1}{12}\,\left(\frac{\lr}{\lso}\right)^2\ .\label{eq:ellipse}
\end{equation}
and at some $k\neq k^*$ for
\begin{equation}
\lr/\lso>2\sqrt{3}\ ,\quad \lnu=0\ .
\end{equation}
\end{subequations}
The contour~(\ref{eq:ellipse}) separates the topological quantum spin Hall phase (inside) and the topologically trivial phase (outside). 
Near $\lambda_\nu=0$ there exists a region in parameter space, where the bands are separated by a finite direct but \emph{no} indirect bandgap; see left plot of \fig{fig:heatmap}.
In this range the different bands are nondegenerate at all points such that the topological invariant is still meaningful; however, the Hamiltonian does \emph{not} describe an insulator any more for these choices of parameters.

\medskip

We now study the \sles\  for this model. According to the discussion in \Sec{ss:sles-topological} we consider the eigenstate 
\begin{equation}
u_{i,j}(\bk)=\left(u_{i,j}^{\rm A\uparrow}(\bk),u_{i,j}^{\rm A\downarrow}(\bk),u_{i,j}^{\rm B\uparrow}(\bk),u_{i,j}^{\rm B\downarrow}(\bk)\right)^t
\end{equation}
at the Dirac point $\bk^*$. For the four bands with energies given in \eq{eq:kmm-ene} the (unnormalized) eigenstates are
\begin{subequations}
\begin{align}
u_{1,j}(\bk^*)&=\left(\kappa_j,0,0,1\right)^t\\
u_{2,j}(\bk^*)&=\left(0,\kappa_j,1,0\right)^t\ ,
\end{align}
\label{eq: states under consideration}
\end{subequations}
with
\begin{equation}
\kappa_j=\frac{-\i}{3\lr}\left(2\lnu+(-1)^j\,\sqrt{4\lnu^2+9\lr^2}\right)\ .
\end{equation}
The energy $\varepsilon_{1,1}$ is always negative [see \eq{eq:kmm-ene}] so according to \eq{eq:projection} the projector $\pp$ on the states~\eqref{eq: states under consideration} is
\begin{equation}
\pp=\begin{pmatrix}
|\kappa_1|^2&0&0&\kappa_1\\
0&|\kappa_1|^2&\kappa_1&0\\
0&\kappa_1^*&1&0\\
\kappa_1^*&0&0&1\ 
\end{pmatrix}.
\end{equation}
For general parameter values, this projector does not have a clear physical interpretation.
However, for $\lr=0$ and $\lnu\neq0$ the constant $\kappa_1$ vanishes  and  $\pp$ becomes the projector on the sublattice B,
\begin{align}
\pp=
\mathrm{diag}(0,0,1,1)\ .
\label{eq: projection on sl B}
\end{align}
In the following, we will use the projection~\eqref{eq: projection on sl B} on sublattice B for \emph{all} choices of parameters. 

The \sles\ is now given by two eigenvalues $\lambda_{1,2}$ of $1/2-\pp\Pi(\bk) \pp$. The right panel of \fig{fig:heatmap} shows choices of parameters $\lnu$ and $\lr$ where the \sles\ fills the whole interval $[-1/2,1/2]$. This exactly coincides with the topological phase inside the ellipse defined in \eq{eq:ellipse}.

This feature can be understood from an inversion of the energy bands $\varepsilon_{1,2}$ and $\varepsilon_{2,1}$: 
While $\varepsilon_{1,1}$ is always negative and $\varepsilon_{2,2}$ is always positive, $\varepsilon_{1,2}$ and $\varepsilon_{2,1}$ change signs exactly at the phase transition~(\ref{eq:ellipse}). For dominant spin-orbit coupling (inside the topological phase) we have $\varepsilon_{1,2}<0$ and $\varepsilon_{2,1}>0$, while for dominant Rashba coupling or sublattice potential (in the trivial phase) we find $\varepsilon_{1,2}>0$ and $\varepsilon_{2,1}<0$. During the phase transition the eigenstates $u_{1,2}(\bk^*)$ and $u_{2,1}(\bk^*)$ change from occupied to empty and from empty to occupied, respectively, changing $\Pi(\bk)$ qualitatively.

In the topological phase the two entanglement eigenvalues are $\lambda_{1}(\bk^*)= 1/2$ and $\lambda_{2}(\bk^*)= -1/2$. As $\pp$ defined in Eq.~\eqref{eq: projection on sl B} is time-reversal invariant, the Kramer's degeneracy at the \trim\ then leads to a gapless \sles. In contrast, in the topologically trivial phase  $\lambda_{1}(\bk^*)=\lambda_2(\bk^*)=0$, such that the \sles\ is now gapped.

Unlike with the case of the (two-band) Chern insulator, the quantum distance (that is, the sum of the two entanglement eigenvalues of the \sles) admits no conclusions about topology of a $\mathbb{Z}_2$ topological insulator. For example, we find a spectral symmetry $\lambda_1=-\lambda_2$ for $\lr, \lnu=0$ and therefore a constant quantum distance $d(\pp,\Pi(\bk))=1$.

\medskip

In order to illustrate the importance of the correct choice of the bipartitioning, we consider a projection on one of the two spin species. Then, the related projector breaks time-reversal symmetry and therefore we do not find any Kramer's degeneracy. In this case the entanglement spectrum does not reveal the topological character of the system.\cite{hsieh-2013} Rather it measures the coupling of the two spin species which is related to the parameter of the Rashba interaction $\lr$.

\section{Conclusion}\label{s:concl}
We studied the relations between the entanglement properties, the topology, and the quantum geometry of noninteracting band insulators. The connections between these quantities are best seen if the entanglement cut preserves the full translational symmetry of the system, that is, in the SLES.
First, we extended the known relation that a topologically nontrivial band structure implies a gapless entanglement spectrum to the SLES for both Chern insulators and $\mathbb{Z}_2$ topological insulators. 
We are thus trading the topological stability guaranteed by the gap in the spectrum of Hamiltonian for  the topological stability of the ``entanglement Fermi surface" of the entanglement Hamiltonian.
Second, we reinterpreted the trace over the entanglement spectrum at a given momentum as a quantum distance associated with this momentum. This allowed us to establish the Fubini-Study metric of a Bloch band as an upper bound to the squared momentum-derivative of the entanglement eigenvalues. 

For this study, we concerned ourselves with the simplest possible case of noninteracting fermionic Hamiltonians with full translational symmetry.
It is imperative to ask how our results can be extended to interacting ``highly entangled" SPT phases and to phases with intrinsic topological order. This includes relating their many-body entanglement spectra to the many-body metric of the ground state which is defined as the function of external control parameters, such as twists in the boundary conditions.

We close by noting that as the quantum metric tensor of Bloch bands is in part experimentally accessible via optical susceptibility~\cite{Souza00} or current noise measurements~\cite{neupert-2013}, the relations that we found provide ways of experimentally obtaining some information about the entanglement spectrum.

\begin{acknowledgments}
We are very grateful to Claudio Chamon for his inspiring comments and critical reading of the manuscript.
We thank Shinsei Ryu, Luiz Santos, Constantin Schrade, and Manfred Sigrist for stimulating discussions.
This work was in part supported by the Swiss National Science Foundation.
\end{acknowledgments}

\bibliography{../Literatur}

\end{document}